\documentclass{emulateapj}
\usepackage{times}

\def\gs{\mathrel{\hbox{\rlap{\hbox{\lower4pt\hbox{$\sim$}}}\hbox{$>$}}}}
\def\ls{\mathrel{\hbox{\rlap{\hbox{\lower4pt\hbox{$\sim$}}}\hbox{$<$}}}}

\def\xmm{{\it XMM-Newton}}
\def\chandra{{\it Chandra}}
\def\swift{{\it Swift}}

\def\integral{{\it Integral}}
\def\hipparcos{{\it HIPPARCOS}}
\def\rosat{{\it ROSAT}}
\def\cobe{{\it COBE}}
\def\iras{{\it IRAS}}
\def\grb{{GRB~050724}}
\def\et{{et al.\ }}

\def\arcs{{\hbox{$^{\prime\prime}$}}}
\def\degg{^{\circ}}

\begin{document}

\submitted{Accepted 2005 November 02}
\title{The dust-scattered X-ray halo around \swift\ \grb}
\shorttitle{The X-ray halo around \grb}
\shortauthors{Vaughan \et}
\author{S. Vaughan\altaffilmark{1},
R. Willingale\altaffilmark{1},
P. Romano\altaffilmark{2},
J. P. Osborne\altaffilmark{1}, 
M. R. Goad\altaffilmark{1},
A. P. Beardmore\altaffilmark{1},
D. N. Burrows\altaffilmark{3},
S. Campana\altaffilmark{2},
G. Chincarini\altaffilmark{2,4},
S. Covino\altaffilmark{2},
A. Moretti\altaffilmark{2},
P. T. O'Brien\altaffilmark{1},
K. L. Page\altaffilmark{1},
M. A. Supper\altaffilmark{1},
G. Tagliaferri\altaffilmark{2}
}
\email{sav2@star.le.ac.uk}
\altaffiltext{1}{X-Ray and Observational Astronomy Group, University of
       Leicester, Leicester, LE1 7RH, UK}
\altaffiltext{2}{INAF-Osservatorio Astronomico di Brera, Via Bianchi
  46, 23807 Merate, Italy.}
\altaffiltext{3}{Department of Astronomy \& Astrophysics, Pennsylvania
  State University, 525 Davey Lab, University Park, PA 16802, USA.}
\altaffiltext{4}{Universit\'{a} degli studi di Milano-Bicocca,
  Dipartimento di Fisica, Piazza delle Scienze 3, I-20126 Milan,
  Italy.} 


\begin{abstract}

This paper discusses the X-ray halo around the \swift\ $\gamma$-ray
burst \grb\ ($z=0.258$), detected by the \swift\ X-Ray Telescope.
The halo, which forms a ring around the fading X-ray source,
expands to a radius of $200\arcs$ within $8$~ks of the
burst exactly as expected for small-angle X-ray scattering by 
Galactic dust along the line of sight to a cosmologically distant
GRB. The expansion curve and radial profile of the halo constrain the scattering
dust to be concentrated at a distance of $D = 139 \pm 9$~pc (from Earth)
in a cloud/sheet of thickness $\Delta D < 22$~pc.
The halo was observed only out to scattering angles of 
$200\arcs$, for which the scattering is dominated by the
largest grains, with a maximum size estimated to be $a_{\rm max}
\approx 0.4-0.5~\mu$m. 
The scattering-to-extinction ratio was estimated to be $\tau_{\rm
  scat}/A_{\rm V} \gs 0.022$; this is a lower limit to the true value because
contribution from smaller 
grains, which scatter to larger angles, was not directly observed.
The line-of-sight to the GRB passes close to the Ophiuchus
molecular cloud complex, which provides a plausible site
for the scattering dust.
\end{abstract}

\keywords{ gamma rays: bursts --- X-rays: general --- Galaxy:
structure --- ISM:dust}


\section{Introduction}
\label{sect:intro}

Small-angle scattering of X-rays by dust grains can produce a
`halo' around a distant X-ray source, with a radial intensity
distribution that depends on the dust properties and location
(Overbeck 1965; Hayakawa 1970; Tr\"umper \& Sch\"ofelder 1973).
This
effect was first detected by Rolf (1983) using {\it Einstein}
observations of the bright X-ray binary GX~$339-4$, and has subsequently
been observed around several other bright Galactic X-ray sources
(e.g. Catura 1983; Predehl \& Schmitt 1995; Predehl \et 2000).

Gamma-ray bursts (GRBs) produce high X-ray fluxes for short periods,
typically $\ls 1000$~s. Viewed through a substantial column of 
Galactic dust, these impulsive X-ray events may produce halos that
appear to expand on the sky because X-rays scattered at larger angles 
travel a slightly longer path length to the
observer and suffer an increased time delay.  
This time delay, between the direct and scattered light, means the 
later observations of the scattered light provide a view of the GRB
X-ray emission at earlier times and can, in principle,
provide details of the  location, spatial distribution and properties
of the dust (e.g. Alcock \&
Hatchett 1978; Klose 1994; Miralda-Escud\'e 1999; Draine \& Bond 2004).
In particular, assuming the GRB to lie at a cosmological distance,
the distance to the scattering dust can be measured from the
radial expansion of the halo, because the halo size 
increases as $\theta^2 = 2 \tau c / D$, where $\tau = t - t_{\rm X}$
is the time since the pulse of illuminating X-rays and $D$ is the
distance between the dust and observer. 
Observations of GRB dust haloes can thereby provide very
accurate distances to Galactic structures, as first demonstrated
by Vaughan \et (2004) using
an \xmm\ observation of the \integral\ GRB~031203.

This paper discusses the X-ray halo around \grb, 
the first halo discovered by \swift\ (Romano \et 2005) and only the
second ever observed around a GRB. 
The rest of the paper is organised
as follows:
Section~\ref{sect:data} discusses the \swift\ observations and
basic data analysis;
Section~\ref{sect:burst} describes the X-ray properties of the
burst counterpart during the first few ks of the observation;
Section~\ref{sect:halo} discusses the detailed 
analysis of the X-ray halo images; 
Section~\ref{sect:rass-iras} uses the archival \rosat\ and \iras\ data
to probe the ISM in the direction of the GRB; 
and finally the implications of the
results of this analysis are discussed in Section~\ref{sect:disco}.
Throughout this paper the quoted errors correspond to $90$\%
confidence 
regions unless stated otherwise.


\begin{figure*}
\centering
\rotatebox{270}{
\epsscale{0.4}
\plotone{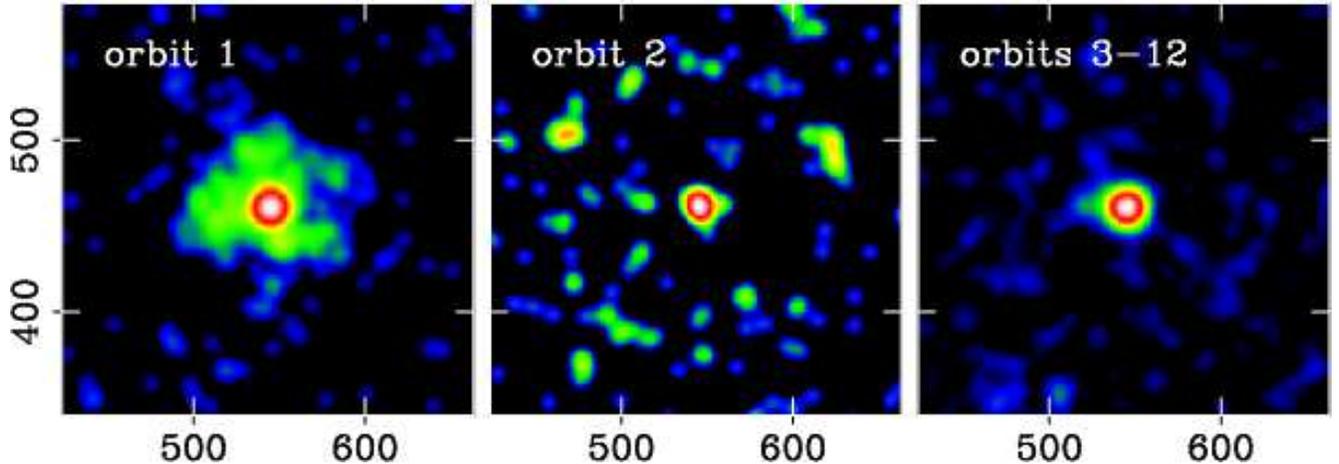}
}
\caption{
$0.2-5$~keV XRT images of \grb\ taken in three time intervals
covering the first orbit ($343-2243$~s after $T_0$), 
second orbit ($6068-8127$~s after $T_0$) and later orbits of \swift,
plotted in detector coordinates.
The images were
smoothed using a Gaussian of width $5$ pixels ($\sigma$).
The X-ray counterpart of \grb\ is
in the centre of each frame. During orbit $1$
the halo expands from a radius of $20$ to $44$ pixels,
during orbit $2$ it is visible as a faint ring 
with a radius of $\sim 85$ pixels,
and in later orbits the halo is not detected.
\label{fig:image}}
\end{figure*}

\section{Observations and data analysis}
\label{sect:data}

At 2005-07-24 12:34:09 UT the \swift\ Burst Alert Telescope (BAT;
Barthelmy 2004) triggered and located \grb\ (Covino \et 2005).  
The prompt BAT light curve showed a  relatively short-lived spike,
with a FWHM of $ \approx 0.25$~s (Covino \et 2005; Krimm \et 2005),
plus an extended low-flux tail lasting for at least $\gs 200$~s.
As discussed by Barthelmy \et (2005), this burst would have
been classified a short burst if observed by BATSE.
The total fluence was $\approx 6.3 \pm
1.0 \times 10^{-7}$~erg cm$^{-2}$ (Krimm \et 2005) over the
$15-350$~keV band. The spectrum evolves from relatively
hard during the $1$~s peak (photon index of $\Gamma = 1.71 \pm 0.16$)
to much softer ($\Gamma = 2.5 \pm 0.2$) at $50-150$~s after the
trigger. 

The spacecraft slewed immediately to the BAT on-board calculated
position; the \swift\ X-ray 
telescope (XRT; Burrows \et 2004, 2005a) began observations at 12:35:22.9 UT
(in automatic state), only $74$~s after the BAT trigger, and detected a
rapidly fading, uncatalogued X-ray source at (J2000) RA=$16^{\rm h} 24^{\rm
m} 44.4^{\rm s}$, Dec$=-27\degg$~$32$\arcmin~$28\arcsec$ (with a
$90$\% confidence radius of $6\arcs$; Barthelmy \et
2005). This position is
close to the Galactic plane ($l=350.37\degg, b=+15.10\degg$);
the expected column density of neutral Galactic gas along this
line-of-sight (LoS) is $N_{\rm H} = 1.46 \times
10^{21}$  cm$^{-2}$ (as measured from $21$~cm maps; Dickey \&
Lockman 1990), and the reddening $E(B-V) = 0.59$ (from the IR dust maps of
Schlegel, Finkbeiner \& Davis 1998).

Follow-up observations revealed
a variable optical and radio source within the XRT error circle
($4.3$~\arcsec\ from the 
XRT position; Gal-Yam \et 2005; Soderberg \et 2005). 
A $50$~ks \chandra\ observation 
further refined the position of the fading
X-ray source and showed it to be coincident with the optical and
radio transient (Burrows \et 2005b).
Prochaska \et (2005) identified the host galaxy of the
transient to be a massive early-type galaxy at 
a redshift of $z=0.258 \pm 0.002$ (from Ca H+K
and G-band absorption). See also Berger \et (2005) for
a discussion of the host galaxy optical spectrum.

The XRT collected data in Windowed-Timing (WT) mode from $79-342$~s
post-burst (WT mode allows imaging in one spatial dimension only),
after which time the data were taken in Photon Counting 
(PC) mode. WT mode allows imaging in one spatial dimension only,
whereas PC mode produces two-dimensional images,
see Hill \et (2004)  for a full description of the XRT
operating modes.  The XRT data were processed by the \swift\ Data
Center at NASA/Goddard Space Flight Center (GSFC) to level $1$ data
products (calibrated, quality flagged event lists). These were further
processed with {\tt xrtpipeline v0.8.8} into
level $2$ data products. In the subsequent analysis only event grades
$0-2$ were used for the WT mode 
data and only grades $0-12$ were used for the PC mode data.


\section{Early X-ray emission from \grb}
\label{sect:burst}

Spectra and light curves were extracted from the WT data using
a $40$ pixel ($94$\arcsec) wide box to define the source and background
regions. For the PC mode data, source counts were accumulated from within
a circle of radius $25$ pixels ($59$\arcsec), and background
data were accumulated within an annulus having inner and
outer radii of $60$ and $100$ pixels ($141-236\arcs$). 
The appropriate ancillary response matrices were generated using {\tt
  xrtmkarf v0.4.14}.  

\begin{figure}
\centering
\rotatebox{270}{
\epsscale{0.80}
\plotone{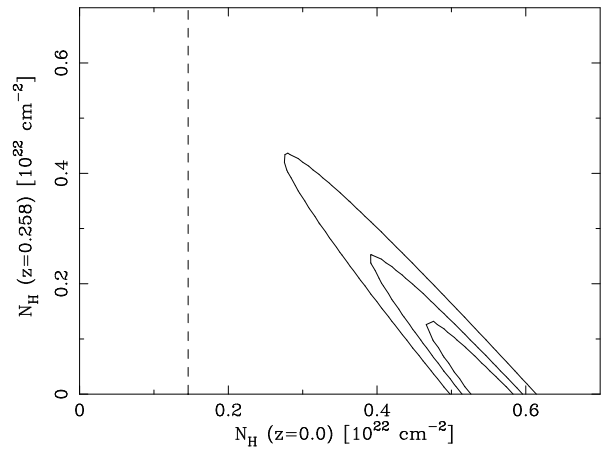}
}
\caption{
$\Delta \chi^2$ contours for absorption in the host Galaxy
($z=0.258$) against Galactic absorption ($z=0.0$). 
The contours correspond to $\Delta \chi^2 =  2.30, ~4.61$ and $9.21$,
respectively. The dotted line indicates the expected $N_{\rm H}$ based
on the $21$~cm measurements of Dickey \& Lockman (1990).
\label{fig:nh}}
\end{figure}

The spectrum of the WT data ($79-342$~s post-burst)
was binned such that each spectral bin contained 
at least $20$ counts and fitted over the $0.6-10$~keV range\footnote{
The bandpass for the WT-mode spectral fit was restricted
to energies above $0.6$~keV to avoid calibration uncertainties at
lower energies.} using {\tt XSPEC} (Arnaud 1996).
The spectral model comprised a power law  
continuum modified by two neutral absorbers (modelled using the {\tt
  TBabs} code 
of Wilms, Allen \& McCray 2000), one at $z=0.0$ and the other at $z=0.258$,
corresponding to neutral interstellar gas in our Galaxy and intrinsic to 
the GRB host galaxy, respectively.
Allowing both column densities to be free parameters,
this simple model gave
a very good fit ($\chi^2 = 355.89$ for $334$ degrees of freedom, dof)
with best-fitting parameters as follows:
$\Gamma = 1.94 \pm 0.05$,
$N_{\rm H} (z=0.0) = 5.9_{-1.5}^{+0.3} \times 10^{21}$ 
cm$^{-2}$ and $N_{\rm H} (z=0.258) < 2.4 \times 10^{21}$ 
cm$^{-2}$.
The Galactic column density inferred from the X-ray spectrum
exceeds that expected based on the $21$~cm
measurements (see Section~\ref{sect:data}), 
by $4.4_{-1.5}^{+0.3} \times 10^{21}$ cm$^{-2}$,
and yet there is no clear excess absorption intrinsic to the host galaxy. 
Figure~\ref{fig:nh} shows the $\Delta \chi^2$ contours for the two
absorption column densities.

\section{X-ray halo analysis}
\label{sect:halo}

\subsection{Expansion curve}

\begin{figure}
\centering
\rotatebox{270}{
\epsscale{0.80}
\plotone{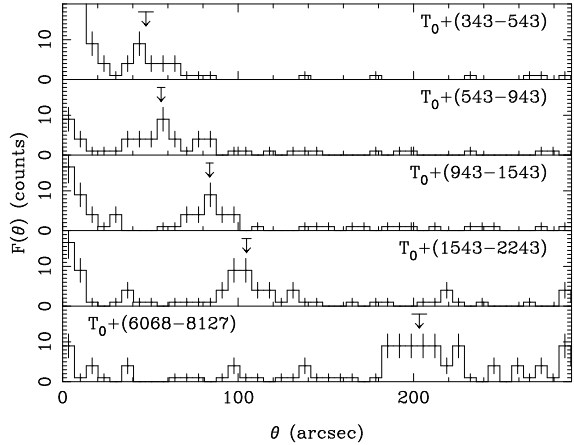}
}
\caption{
Radial profile of counts about the \grb\ afterglow from five
time intervals. The times are marked in each frame.
The arrows mark the maximum likelihood
estimate of the halo radius, $\theta$, derived from fitting these
profiles, and the horizontal bar indicates the $68.3$\% ($1\sigma$)
confidence interval on the radius.
\label{fig:profiles}}
\end{figure}

Figure~\ref{fig:image} shows the XRT (PC mode) images 
from the first,
second and later orbits of \swift, over the $0.2-5$~keV energy
range (only events with energies $0.2-5$~keV were used since the halo
has a soft spectrum, as discussed in Section~\ref{sect:halo-spec}, and
was not detected above $5$~keV).
The first orbit showed
extended emission around the point-like GRB which 
was dispersed in the second orbit and was difficult to 
detect in later orbits. 
In order to better quantify this extension, images
were accumulated in five non-overlapping time intervals:
$343-543$, $543-943$, $943-1543$ and $1543-2243$~s from 
the first orbit and $6068-8127$~s from the second orbit.
 
\begin{figure}
\centering
\rotatebox{270}{
\epsscale{0.80}
\plotone{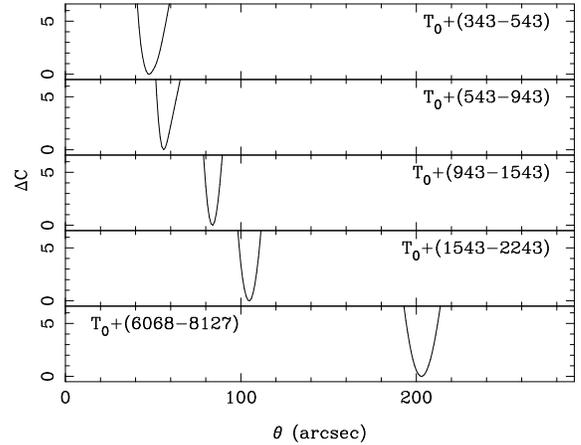}
}
\caption{
Change in $C$-statistic as a function of halo position.
The minimum of each curve represents the maximum likelihood
estimate of the halo radius.
Nominal $68.3$ and $90$\% confidence regions can be defined
by the intervals bounded by $\Delta C = 1.0$ and $2.706$, respectively.
\label{fig:c-stat}}
\end{figure}

Radial profiles were
calculated for each image by accumulating the counts in annuli
centred on the GRB
(see Figure~\ref{fig:profiles}).  
The expanding halo was revealed by the small
but significant localised excess of counts, beyond the 
unresolved GRB\footnote{The  Point Spread Function (PSF) of
the XRT has a half-energy radius of $\approx 8.8\arcs$,
or FWHM$\approx 9.6\arcs$ at $E\approx 1.5$~keV (Moretti \et 2005;
Burrows \et 2004)}. 
The radius of the halo as a function of 
time, $\theta (t)$, was estimated by  fitting each radial profile
(using {\tt XSPEC}) with a model comprising the
PSF of the source (a King profile with parameters taken from ground
based calibration data; Moretti \et 2004, 2005) plus a constant background
level (per pixel), integrated over the annular bin.  A Gaussian was
added to represent the profile of the halo, fixing the width
to be equal to $10$\% of the radius: $\sigma_{\theta} = 0.1 \theta$
(see Section~\ref{sect:comoving}).
The $C$-statistic was used to find the maximum likelihood model
parameters, appropriate for the case of few counts per bin
($C=-2\log \mathcal{L}$, where $\mathcal{L}$ is the likelihood; Cash
1979). For each profile, the
$C$-statistic was minimised and the maximum likelihood position of the
Gaussian was used as the radius of the halo in that image. The $90$\% 
confidence region for the radius was estimated using the $\Delta
C$ values (see Figure~\ref{fig:c-stat}).

\begin{figure}
\centering
\rotatebox{270}{
\epsscale{0.80}
\plotone{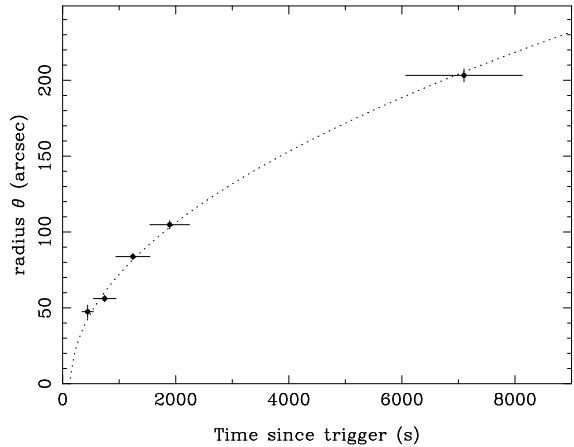}
}
\caption{
Expansion of the ring around \grb\ with time.  The radii
were measured by fitting the local maximum in the radial
profiles (Fig.~\ref{fig:profiles}). The expansion was
fitted with the function $\theta(t) = N (t-t_{\rm X})^{1/2}$
(marked with the dotted line). 
\label{fig:expansion}}
\end{figure}

Figure~\ref{fig:expansion} shows the radius of the halo in each of the
five time bins, as measured using the radial profiles.
These data were fitted with the function
$\theta(t) = N (t-t_{\rm X})^{1/2}$, which describes the radial expansion of
a dust-scattered halo following a (short) X-ray pulse at time $t_{\rm X}$
(e.g. Tr\"umper \& Sch\"ofelder 1973; Klose 1994; Miralda-Escud\'e
1999; Draine \& Bond 2004).
Strictly speaking, this assumes a very narrow pulse of X-rays
and an intrinsically narrow dust scattering screen.
The value of $t_{\rm X}$  corresponds to the typical 
arrival time of direct (un-scattered) photons, which will be 
after $T_0$ since the \grb\ remained X-ray bright from 
$T_0$ until $\approx T_0 + 300$~s, meaning almost all the direct X-rays
actually arrived after $T_0$.
The best-fitting values of the free parameters
were: $N = 2.46 \pm 0.08$ (with $\theta$ measured in arcsec)
and $t_{\rm X} = 145_{-74}^{+65}$~s after $T_0$.
This model gave a good fit, with $\chi^2 = 6.70$ for $4$ dof.
In order to check whether the best-fitting value of $t_{\rm X}$ is
plausible, the mean arrival time for the direct X-rays 
was estimated using the X-ray light curve from the joint BAT-XRT data
(Figure~3 of Barthelmy \et 2005):
the flux-weighted photon arrival time\footnote{
The flux-weighted photon arrival time was calculated
as $\langle t_{\rm X} \rangle = \int f_{\rm X}(t) t ~ dt / \int f_{\rm X}(t)
~ dt$ where $f_{\rm X}(t)$ is the X-ray light curve.
} over this light curve was $T_0 +
86$~s, consistent with the best-fitting $t_{\rm X}$ from
the radial expansion curve.

The normalisation, $N$, describes how fast the
halo expands and is uniquely determined by the distance to the
dust\footnote{
The halo shape is affected by the geometry of the
scattering region, and it may not be circular and centred on the
GRB image if the scatterer is in the form of a plane-parallel slab
not perpendicular to the line-of-sight, or is curved (see Tylenda 2004
for a discussion).
However, because of the large ratio of dust distance to halo
size, only the most pathological geometries (e.g. slab included
at $>89\degg$) will produce any observable effect.
}. Assuming the GRB to be at a cosmological distance
$N=(2c/D)^{1/2}$.
The best-fitting normalisation predicts a distance to the
scattering medium of $D = 139 \pm 9$~pc, where the $90$\% confidence
region was calculated with $t_{\rm X}$ as a free parameter\footnote{
This distance estimate is smaller than that given in 
Romano \et (2005) because that earlier analysis assumed
$t_{\rm X} = T_0$, an assumption not made in the present analysis.}

This distance measurement is quite robust to the choice of
function used to parameterise the halo radial profile.
This was demonstrated by repeating the radial profile fitting 
using different functions to model the halo profile.
For each one the halo expansion curve was measured and the
distance calculated from its best-fitting normalisation.
A Gaussian of fixed width ($\sigma=5\arcs$) gave $D=145\pm8$~pc, a
Lorentzian of 
fixed width ($\Gamma = 10\arcs$) gave $D=142\pm8$~pc and a King model
as used to model the PSF (see above) gave $D=144\pm6$~pc. 
These different estimates are all well within the confidence region
of the original estimate, demonstrating that any systematic errors
introduced by the choice of radial profile model are smaller than
the quoted statistical error.

\subsection{Co-moving radial profile}
\label{sect:comoving}

\begin{figure}
\centering
\rotatebox{270}{
\epsscale{0.80}
\plotone{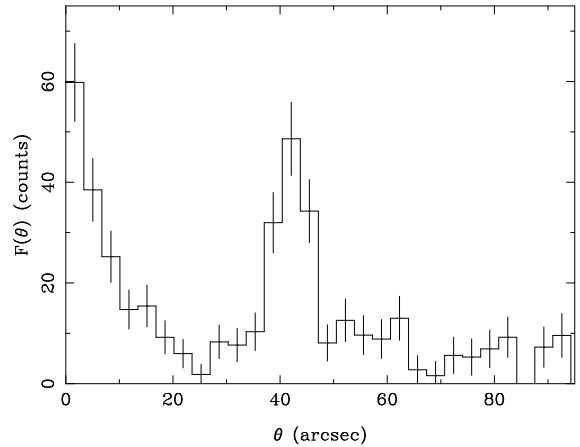}
}
\caption{
Co-moving radial profile around \grb.
This shows the counts accumulated in annuli around the GRB
summed over the five time intervals of Figure~\ref{fig:profiles},
after correcting for the expansion of the halo
(matching the radii to those from the first image).
The halo occurs at $\theta \approx 45$~arcsec from the GRB ($\theta = 0$) 
and is a relatively narrow ($\sigma_{\theta} /
\theta \approx 0.07$) structure. 
\label{fig:comoving}}
\end{figure}

The radial profile of the halo was examined in more detail 
by summing the five radial profiles (Figure~\ref{fig:profiles}), after
adjusting the 
size of the radial bins to correct for the expansion of the
halo. 
A constant background level, estimated from the mean level at large
radii, was subtracted from each profile before combining them.
The result is shown in Figure~\ref{fig:comoving}. 
This shows that the average radial profile of the halo, after correcting for 
its expansion, was reasonably narrow.
Fitting this profile using a Gaussian to model the halo
gave a relative width of ${\rm FWHM}(\theta) / \theta \approx
0.16-0.23$ ($90$\% confidence limit).

The observed radial width of the halo is a combination of four
terms:
($i$) the time profile of the illuminating X-ray pulse;
($ii$) the radial distribution of the dust along the line of sight, 
($iii$) the PSF of the XRT,
and ($iv$) the radial expansion of the halo over the length of an exposure.
Term ($i$) is negligible since the time duration of the initial pulse
is short compared to the times (since burst) of the halo images
($343-8127$~s). 
The XRT PSF (with FWHM of $9.6\arcs$) gives rise to a fractional width
of $20$\% in the first image to $5$\% in the last, meaning  term
($iii$) must contribute significantly to the measured width.
The co-moving radial profile is dominated by counts
from later images (with longer exposures), 
during which the fractional increase in the halo
radius -- term ($iv$) -- was $\sim 15$\%. 
The measured width of the halo may therefore be explained entirely by
terms ($iii$) and ($iv$), implying the remaining term, the radial
thickness of the dust, must be small.
A very conservative limit on the 
physical thickness of
the dust cloud along this line 
of sight is $\Delta D \approx D \times {\rm
  FWHM}(\theta)/\theta < 22$~pc. 
Most likely the dust is confined in a much narrower region.

\subsection{Halo spectrum}
\label{sect:halo-spec}

\begin{figure}
\centering
\rotatebox{270}{
\epsscale{0.80}
\plotone{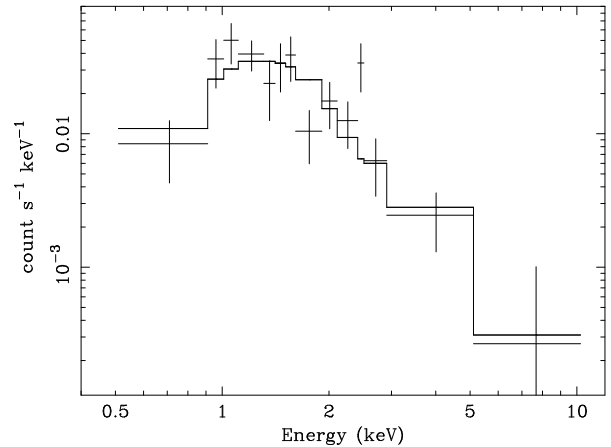}
}
\caption{
XRT spectrum of the halo during the first \swift\ orbit.
Crosses mark the data and the histogram marks the
best-fitting absorbed power law model.
The data were binned for illustration purposes only.
\label{fig:halo-spec}}
\end{figure}

The spectrum of the halo was extracted
from the first orbit of data using  
an annular extraction region with inner and outer radii of $20$ and $45$
pixels ($47-106$~arcsec), respectively. A background spectrum 
was extracted from an annulus with radii of $60$ and $100$
pixels and an appropriate ancillary response file was generated. 
Due to the small number of counts ($\approx 100$) the
spectrum was fitted using the $C$-statistic, rather than
binning to $20$ counts per bin and fitting using the
$\chi^2$ statistic, which would result in very few bins.
The spectral model used was an absorbed power law, with the
neutral absorption column density fixed at the value obtained from the 
early XRT spectrum ($N_{\rm H} = 5.9\times 10^{21}$~cm$^{-2}$;
section~\ref{sect:burst}). The data are shown in Figure~\ref{fig:halo-spec}.
The best fitting photon index was $\Gamma =
2.7_{-0.3}^{+0.5}$, significantly steeper than the GRB X-ray spectrum ($\Gamma
= 1.94 \pm 0.05$), as expected due to the strong energy dependence of the
dust scattering cross section.
The model gave a good fit, with a rejection probability of
only $p=0.31$ estimated using $10^3$ Monte Carlo simulations.
The mean photon energy in the binned, background subtracted halo 
spectrum was $\langle E \rangle = 1.54$~keV.

\subsection{Halo flux as a function of angle}
\label{sect:decay}

\begin{figure}
\centering
\rotatebox{270}{
\epsscale{0.80}
\plotone{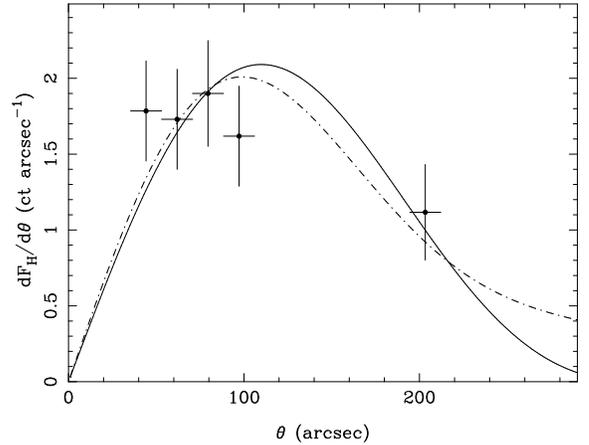}
}
\caption{
Halo flux as a function of scattering angle.
The crosses show the background-subtracted counts
per (equal size) radial bin over which the halo was detected.
The solid curve marks the best-fitting theoretical scattering
curve $dF_{\rm H}/d\theta$ (Equation~\ref{eqn:scat}) assuming 
a single grain size ($a=0.367~\mu$m).
The dot-dashed line shows the best-fitting function assuming a
MRN distribution of grain sizes (with $a_{\rm max}
=0.524~\mu$m). 
\label{fig:decay}}
\end{figure}

The flux decay of the halo was also estimated using the radial profile
of the combined image from orbits $1$ and $2$ (over which the halo is
well-detected). Counts were accumulated in annuli of
width $\Delta \theta = 17.6\arcs$ ($7.5$ pixels) and background
subtracted assuming a constant background level per pixel estimated
from an annulus enclosing  $280-370\arcs$. The sampling of the halo as
a function of angle is incomplete, due to the incomplete time coverage
of the observation, therefore radii over which the halo was not
observed were masked out of the radial profile.  The masking was done
by converting the times of the  observations into radii using the
best-fitting expansion curve (Figure~\ref{fig:expansion}) and
removing all radial bins containing unobserved angles.  The result is
shown in Figure~\ref{fig:decay}.

The total fluence of scattered light (in the halo), $F_{\rm H}$,
should be related to the direct light (i.e. un-scattered light from the
GRB), $F_{\rm X}$, by:
\begin{equation}
\frac{F_{\rm H}}{F_{\rm H}+F_{\rm X}} = 1 - {\rm e}^{-\tau_{\rm scat}}
\end{equation}
where $\tau_{\rm scat} = \sigma N_{\rm g}$ is the scattering optical
depth, with 
$\sigma$ the total scattering cross-section and $N_{\rm g}$ is the
dust grain column density along the LoS (e.g. equation~14
of Mauche \& Gorenstein 
1986, henceforth MG86). For low  optical depths 
($\tau_{\rm scat} \ll 1$), where multiple scattering is insignificant, 
this can be approximated by 
$F_{\rm H} \approx \tau_{\rm scat} F_{\rm X}$.
The above quantities represent the entire halo
(integrated over all scattering angles) 
but the scattering cross-section is a function of angle which
defines the halo fluence per unit angle.
In the Rayleigh-Gans approximation\footnote{
Smith \& Dwek (1998) demonstrated that the
Rayleigh-Gans theory and the more exact Mie theory
are in close agreement for photons  energies $E\approx 2$~keV and
typical grain sizes. } the differential scattering
cross-section for a spherical grain of size 
$a$ (in units of $\mu$m) is given by:
\begin{equation}
\frac{d\sigma}{d\Omega} = A_E a^6
\left[ \frac{j_1(x)}{x}\right]^2 (1+\cos^2 \theta)
\label{eqn:cross-sect}
\end{equation}
(equation 2.2 of Hayakawa 1970)
where $A_E$ is a normalisation that depends on the energy of X-rays
being scattered, $x=(4 \pi a /\lambda) \sin(\theta/2)$ and
$j_1(x) = (\sin x)/x^2 - (\cos x)/x$ is the spherical Bessel 
function of the first order.
The central core of this function is approximately Gaussian,
leading to a halo with a  characteristic size
$\theta_{\rm rms} \approx 62.4(aE)^{-1}$~arcsec (equation~6 of MG86;
Hayakawa 1970).
Using equation~\ref{eqn:cross-sect}
the scattered fluence in the halo per unit scattering angle is:
\begin{eqnarray}
\frac{dF_{\rm H}}{d\theta} & = & 
A_E a^6 2 \pi \theta \left[ \frac{j_1(x)}{x}\right]^2 (1+\cos^2 \theta)
N_{\rm g} F_{\rm X} \nonumber \\
 & \propto &
a^6 \theta \left[ \frac{j_1(x)}{x}\right]^2 (1+\cos^2 \theta)
\label{eqn:scat}
\end{eqnarray}
which peaks at $\theta \approx \theta_{\rm rms}$.

This function was fitted to the data (solid curve
in Figure~\ref{fig:decay}) yielding a best-fitting value of
$a=0.367~\mu$m (with a $90$\% confidence region of $0.316-0.430~\mu$m),
assuming a typical energy of 
$E\approx1.54$~keV for the scattered X-rays (see
section~\ref{sect:halo-spec}).
The scattering of soft X-rays is
dominated by the largest grains, due to the strong dependence
of the scattering cross-section on grain size (the total scattering
cross section, integrated over all angles is  $\sigma \propto a^4$), and so
this estimate for 
$a$ corresponds to grains close to the maximum grain size. 
This best-fitting model was integrated over all angles to estimate the
total halo fluence of $F_{\rm H} \approx 342\pm34$ counts
(where the $1\sigma$ error was approximated by the $10$\% fractional error
on the total observed halo fluence).

The above calculation is valid for a single grain size
which should be a reasonable approximation given the strong dependence
of the scattering cross-section on $a$.
The presence of smaller grains  will however cause the halo
fluence to fall off less rapidly at large angles and 
this extended `tail' could contain substantial
additional fluence. In order to investigate the consequence of
a continuous grain size distribution, equation~\ref{eqn:scat}
was re-calculated at each angle by integrating over the grain
size distribution $N(a)$. The size distribution was
assumed to  follow that of Mathis, Rumpl \& Nordsieck (1977;
hereafter MRN):
\begin{eqnarray}
N(a) = \left\{
\begin{array}{ll}
N_0 a^{-q} ~~ & a_{\rm min} \le a \le a_{\rm max} \\
 & \\
0 & {\rm elsewhere}
\end{array}
\right.
\end{eqnarray}
The revised function was fitted to the data assuming $a_{\rm min}
\approx 0.04~\mu$m and $q=3.5$, yielding a best-fitting $a_{\rm max} =
0.524~\mu$m (with a $90$\% confidence region $0.426 - 0.790~\mu$m),
this is slightly larger than, but not inconsistent with
other estimates derived from X-ray haloes
(MG86; Predehl \et 1991; Clark 2004). 
The best-fitting model is
also shown in Figure~\ref{fig:decay} (dot-dashed curve). 
The fit is not very sensitive to $a_{\rm min}$\footnote{
Using $a_{\rm min}=0.004~\mu$m or $0.1~\mu$m changed the fit statistic
by only $\Delta \chi^2 = +0.009$ or $+0.038$, respectively.} as
long as $a_{\rm max}/a_{\rm min} \gs 5$ and $q \gs 3.5$.
The total fluence in this model, integrated over all angles,  was
virtually identical to the single grain size model ($\approx 343\pm34$
counts), although the fluence could be higher if there is a greater
contribution from smaller grains scattering to larger angles that
were not directly observed.

\subsection{Scattering optical depth}
\label{sect:tau}

Given the above estimate of the halo fluence, $F_{\rm H}$, it 
is possible to estimate the dust scattering optical depth once
the direct GRB X-ray fluence, $F_{\rm X}$, is known.
This latter quantity can 
be estimated by integrating the combined XRT-BAT light curve of the burst.
Figure~3 of Barthelmy \et (2005) shows the X-ray light curve
of \grb\ from $T_{0}$. This was produced using the XRT observations
after $T_{0}+79$~s, and the $15-25$~keV BAT light curve extrapolated into 
the XRT band based on a spectral fit to the simultaneous BAT and XRT
spectrum. 
During the first $\sim 300$~s after the trigger the
X-ray source was bright; the WT data contain $\sim 1.18 \times 10^4$
source counts ($0.2-10$~keV).
The extrapolated BAT data predict almost exactly the same
number of counts prior to the start of the XRT/WT observation.
Although there will be a systematic error associated with the
extrapolation of the BAT data to lower energies, the tight
agreement between BAT and XRT data in the interval of overlap
suggests this is likely to be small (Barthelmy \et 2005).
Together these give a total of $\approx 2.42\times10^4$ counts.
Combining this with the $F_{\rm H}$ value of section~\ref{sect:decay}
gave $\tau_{\rm scat} \approx F_{\rm H} / F_{\rm X} \approx 0.014$.
Making an approximate correction to an effective scattering optical depth at
$1$~keV gives $\tau_{\rm scat} (1~{\rm keV}) \approx \tau_{\rm scat} (E) E^2 
= 0.034$ (with $E$ the mean scattered photon
energy in keV). 



\section{ISM towards \grb}
\label{sect:rass-iras}

\begin{figure}
\centering
\plotone{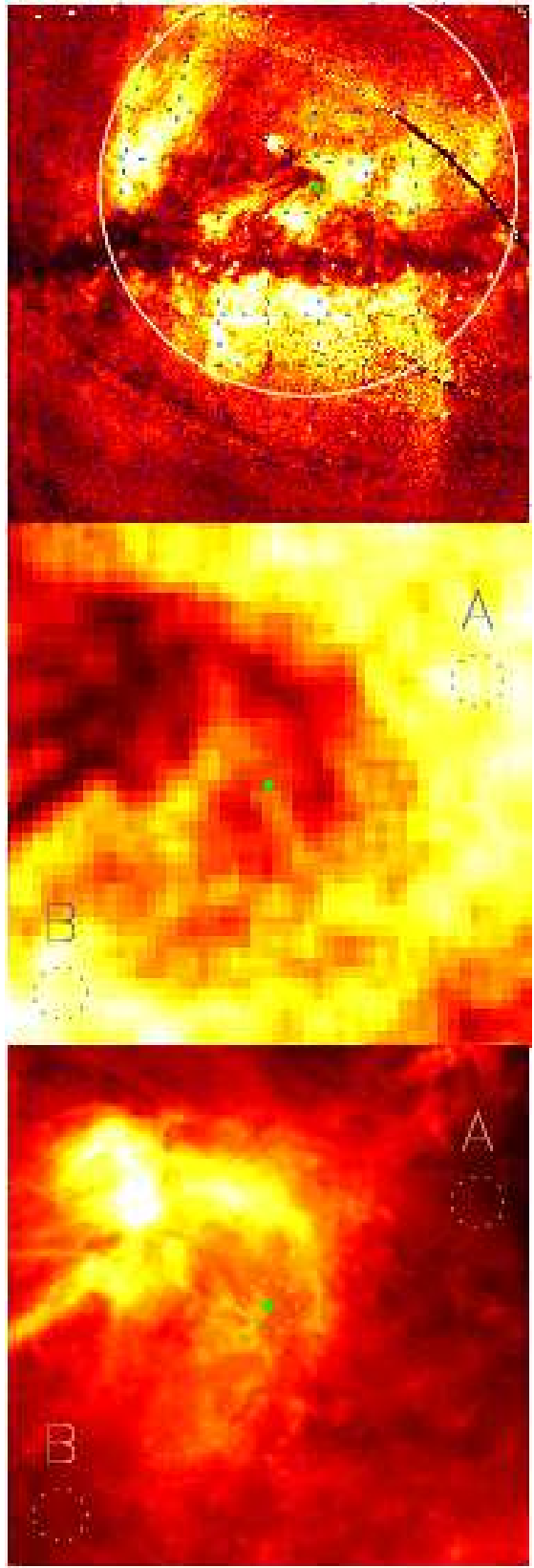}
\caption{
X-ray and IR images of the sky around \grb.
Upper panel: $0.4-1.2$~keV \rosat\ All-Sky Survey (RASS)
map in Galactic coordinates spanning $102\degg$ on a side. 
The position of \grb\ is
marked with the small filled circle, the large circle
marks the extent of the super-bubble centred on the Sco-Cen
OB association.
Middle panel: $10\degg\times10\degg$ close up of the RASS map
around \grb\ (marked with the filled circle).
The dotted circle has a radius of $0.5\degg$, which 
corresponds to $\approx 2.4$~pc at the distance to the
scattering dust.
Lower panel: $100~\mu$m IRAS map on the same scale
as the middle panel. 
Clearly the IR emission and soft X-ray absorption are well
correlated. 
Again the position of the GRB is indicated by
the filled circle.
Positions `A' and `B' mark the off-cloud regions discussed in the
text. 
\label{fig:rass-iras}}
\end{figure}

Archival \rosat\ and \iras\ data were used to 
constrain the ISM in the direction of \grb.
The top two panels of
Figure~\ref{fig:rass-iras} show the soft X-ray  
sky around \grb, derived from the \rosat\ All-Sky Survey (RASS)
data in the $0.4-1.2$~keV band, and the bottom panel shows the \iras\
All-Sky Survey $100~\mu$m map of the region.
These images reveal  structures dominated by the Ophiuchus molecular
cloud complex 
casting the dark `shadows' in the \rosat\ map (middle panel) and
enhanced IR dust emission in the \iras\ map (lower panel).
The main $\rho$ Oph molecular cloud (L1688; Klose 1986) is the
in the bright region in the upper-left part of the \iras\ map.
These data were used to estimate the column density in the
cloud by comparing the column density estimates for the GRB position to
those at locations `A' and
`B,' chosen to be representative of nearby but off-cloud lines-of-sight.

The column densities  were estimated in 
three independent ways, and the results are presented in 
Table~\ref{tab:nh}. The first estimates comes from the 
Dickey \& Lockman (1990) 
$21$~cm maps of $N_{\rm H}$.
This shows little change in $N_{\rm H}$ between the three positions
(see Table~\ref{tab:nh}), and this was confirmed by the $21$~cm map
of the region shown by de Geus \& Burton (1991; see their figure~5).
The column density due to the intervening cloud in the direction of
\grb\ was estimated from the difference between the column density
towards the GRB and the mean of that measured for A and B. This
gave $N_{\rm H}=0.05\pm0.20\times 10^{21}$ cm$^{-2}$ (where the uncertainty is
a very rough estimate based on half the difference between A and B).
In other words there is no excess H~{\sc i} in the direction of \grb\
due to the intervening dust cloud.

The second estimate comes 
from the IR reddening maps from Schlegel \et (1998)
This gave $E(B-V)$ values, based on the \cobe\
temperature and \iras\ intensity at the position of interest, 
of $0.59$, $0.17$ and $0.25$ for the GRB, A and B, respectively.
These correspond to optical estinctions of 
$A_{\rm V} = 2.37$, $0.68$ and $1.01$, respectively, assuming
$R_V \equiv A_V / E(B-V) = 4.0$
as measured by Vrba, Coyne \& Tapia (1993) in the direction of the
$\rho$ Oph cloud. 
The reddening values were
converted to hydrogen column densities assuming  $N_{\rm H}
=4.93\times 10^{21} E(B-V)$ cm$^{-2}$ (Diplas \& Savage 1994). 
In this case the excess column density over the GRB was $N_{\rm
  H}=1.9\pm0.2\times 10^{21}$~cm$^{-2}$.  
However, as noted by Jura (1980), Vuong \et (2003) and others, the
dust-to-gas conversion factors in the direction of the Oph cloud
are uncertain due to the complexity of modelling the cloud medium.

The third estimate was derived from RASS spectra extracted\footnote{ 
\rosat\ count rates in different channels were extracted
using the HEASARC ``X-ray background tool'' available via
the following address: {\tt http://heasarc.gsfc.nasa.gov/}
} using $0.5\degg$ radius regions centred on the
positions of the GRB, A and B. 
These were fitted with the
model developed by Willingale \et (2003) but, due to the
complexity of the model and low resolution of the data, 
most parameters were fixed at the values
found from an analysis of nearby \xmm\ fields (M. A. Supper, in prep.).
The two parameters allowed to vary in the fit were the 
foreground absorption column density and the emission
measure of the ``Loop I super-bubble'' (indicated by the large circle
in the top panel of Figure~\ref{fig:rass-iras}).
The excess column density towards \grb\ was estimated, by comparison
with A and B, at $N_{\rm H}=2.7\pm0.4\times 10^{21}$~cm$^{-2}$. Given the
systematic errors involved in modelling the complex X-ray spectrum,
and converting from dust reddening factor to equivalent hydrogen
column density, the X-ray and IR derived estimates are
in reasonable agreement.

\begin{table}
\caption{Compilation of Galactic column density estimates}
\begin{center}
\begin{tabular}{lllll}
\hline
Technique & Reference & $N_{\rm H}$ [GRB] & $N_{\rm H}$ [A] & $N_{\rm H}$ [B] \\
          &           & ($10^{21}$ cm$^{-2}$) & ($10^{21}$ cm$^{-2}$) & ($10^{21}$ cm$^{-2}$) \\
\hline
$21$~cm   & $1$       & $1.46$ & $1.22$ & $1.61$ \\
IR map    & $2$       & $2.92$ & $0.83$ & $1.24$ \\
RASS      & $3$       & $3.57$ & $0.52$ & $1.28$ \\
\hline
\end{tabular}
\label{tab:nh}
\end{center}

REFERENCES: ($1$) Dickey \& Lockman (1990), 
($2$) Schlegel \et (1998), 
($3$) this work
\end{table}


\section{Discussion}
\label{sect:disco}

The X-ray halo around \grb, as observed by the \swift\ XRT, has
provided accurate information on the Galactic dust
distribution in this direction. 
In particular, the narrow halo must be caused by a concentration
of dust at a distance of $D=139\pm9$~pc from Earth, in a cloud or sheet
with a thickness $\Delta D < 22$~pc. 
The radial profile of the halo constrains the size of the 
largest grains to be $a_{\rm max} \sim 0.4-0.5~\mu$m, but
the contribution from smaller grains is unconstrained because
the halo was not detected at scattering angles larger than 
$\theta \approx 200\arcs$. The estimated value of $a_{\rm max}$ is
higher than the often used value of $0.25~\mu$m (MRN; see also Predahl
\& Schmitt 1995), although this is not without precedent as Jura
(1980) previously noted that the non-standard optical extinction
curves in the direction of $\rho$ Ophiuchus may signify the presence of
unusually large grains. 

The LoS to \grb\ includes the Upper Scorpius subgroup of the
Scorpius-Centaurus OB association at 
a mean distance of $145\pm2$ pc (derived from \hipparcos\ data; de
Zeeuw \et 1999). There is considerable structure in the extinction 
maps around this region,
including material belonging to the Ophiuchus molecular cloud complex
(de Geus 1992). Knude \& H\o g (1998)
detected a sharp rise in the reddening at $120-150$~pc
in the direction of the $\rho$ Oph star forming region, 
which is consistent with the distance measured for the
dust responsible for the X-ray halo. If the X-ray scattering
dust is associated with part of the Oph molecular cloud
complex, this gives the most accurate measurements to date for the
distance to the cloud and its physical thickness.

The excess optical extinction in the direction of the \grb,
compared to the mean of the off-cloud positions `A' and `B', 
was $A_V=1.5$ mag (Section~\ref{sect:rass-iras}).
In combination with the estimated halo scattering optical depth 
(Section~\ref{sect:tau}) this gives $\tau_{\rm scat}/A_V \approx 0.022$
at $1$~keV. This is lower than the value of $0.056$ 
estimated by Predehl \& Schmitt (1995) using 
\rosat\ observations of bright Galactic X-ray sources,
and a factor of a few lower than the
value derived from the model of Draine (2003).
However, the halo around \grb\ was only observed at 
small angles ($\theta \ls 200\arcs$) and so 
the flux contribution from smaller grains is largely unconstrained, meaning
that the true halo flux, and hence $\tau_{\rm scat}$, is probably larger. 

As shown by Figure~\ref{fig:rass-iras}, the X-ray absorption and IR
dust emission are well-correlated, demonstrating they are caused by
the same medium. By contrast the $21$~cm map shows little correlation
with these images (Figure~5 of de Geus \& Burton 1991), suggesting the
cloud medium has a comparatively low density of atomic H~{\sc i}. 
The most obvious explanation is that the hydrogen is
molecular, although there is no obvious CO 
emission from this location in figure~3 of  de Geus, Bronfman \& 
Thaddeus (1990). The Ly-$\alpha$ map of the
region (see figure~13 of de  Geus \et 1990) shows
two nearby H~{\sc ii} regions,  known as S9 and RCW 129, around the
stars $\sigma$ Sco (B1 III) and $\tau$ Sco (B0 V), respectively. It is
therefore also plausible that some fraction of the atomic hydrogen is
ionised by these nearby young stars (with distances of $\sim 140$~pc;
Shull \& van Steenberg 1985).  The total Galactic hydrogen
column density along the LoS to \grb\ is therefore the sum of that
revealed by the $21$~cm H~{\sc i} maps and that inferred from the
molecular cloud (mostly H$_{2}$ and/or H~{\sc ii}).   The $21$~cm
measurement of the total H~{\sc i} column density gave $1.5\times
10^{21}$~cm$^{-2}$, whereas the column density  in the dust cloud was
measured at $1.9\pm0.2\times 10^{21}$~cm$^{-2}$ from the excess IR
emission and  $2.7\pm0.4\times 10^{21}$~cm$^{-2}$ from the excess soft
X-ray absorption (Section~\ref{sect:rass-iras}).  The total hydrogen
column density is therefore in the range $3.4-4.2\times
10^{21}$~cm$^{-2}$, slightly lower than  that measured from the GRB
X-ray spectrum ($4.4-6.2\times 10^{21}$~cm$^{-2}$;
Section~\ref{sect:burst}). It is of course possible that this extra
X-ray absorption along the LoS to \grb\ ($\Delta N_{\rm H} \sim
1.5\times 10^{21}$~cm$^{-2}$) is caused by either a modest   column of
molecular or ionised gas on the far side of
the Sco-Cen super-bubble, or due to cold gas in the GRB host galaxy,
the formal constraint on which was $N_{\rm H} < 2.4\times 10^{21}$
cm$^{-2}$ (Section~\ref{sect:burst}).

{\it Note added in proof:} Since the completion of this paper
Tiengo \& Mereghetti (2005) 
have announced the detection of another dust-scattered X-ray halo
around a GRB. This brings the total number of known GRB
X-ray halos to three.


\acknowledgements

SV, MRG, KP, APB, JPO and MAS gratefully acknowledge funding through
the PPARC, UK.  This work is supported at Pennsylvania State
University (PSU) by NASA contract NAS5-00136, at the University of
Leicester (UL) by the Particle Physics and Astronomy Research Council
on grant number PPA/Z/S/2003/00507, and at the
Osservatorio Astronomico di Brera (OAB) by funding from ASI on grant
number I/R/039/04. We gratefully acknowledge the contributions of
dozens of members of the XRT team at PSU, UL, OAB, GSFC, ASI Science
Data Center, and our subcontractors, who helped make this instrument
possible. We thank an anonymous referee for a thoughtful report.


\end{document}